\documentclass[twocolumn,showpacs,preprintnumbers,amsmath,amssymb]{revtex4}
\usepackage{graphicx}
\usepackage{dcolumn}
\usepackage{bm}
\begin{document}
\title{Wetting and capillary nematization of binary hard-platelet and hard-rod fluids}
\author{L. Harnau}
\email{harnau@fluids.mpi-stuttgart.mpg.de}
\author{S. Dietrich}
\affiliation{Max-Planck-Institut f\"ur Metallforschung,
              Heisenbergstrasse 1, D-70569 Stuttgart, Germany,\\
        and Institut f\"ur Theoretische und Angewandte Physik,
     Universit\"at Stuttgart,\\Pfaffenwaldring 57, D-70569 Stuttgart, Germany
}
\date{\today}
\begin{abstract}
Density-functional theory is used to investigate the phase behavior of colloidal
binary hard-platelet and hard-rod fluids near a single hard wall or confined 
in a slit pore. The Zwanzig model, in which the orientations of the particles of 
rectangular shape are restricted to three orthogonal orientations, 
is analyzed by numerical minimization of the grand potential functional. 
The density and orientational profiles as well as the surface contributions to 
the grand potential are determined. The calculations exhibit a wall-induced 
continuous surface transition from uniaxial to biaxial symmetry for 
the hard-rod fluid. Complete wetting of the wall -- isotropic liquid interface by a
biaxial nematic film for rods and a uniaxial nematic film for platelets is found.
For the fluids confined by two parallel hard walls we determine 
a first-order capillary nematization transition for large slit widths, 
which terminates in a capillary critical point upon decreasing the slit width. 
\end{abstract}
\maketitle
\section{Introduction}
Many complex fluids used in industry or in soft condensed-matter
laboratories consist of non-spherical colloidal particles \cite{mait:00}.
In particular suspensions of hard platelike colloidal particles have recently received 
experimental \cite{brow:98,kooi:98,brow:99,kooi:01,kroo:01a}
and theoretical attention \cite{bate:99,gali:00,harn:01a,wens:01,wens:02,harn:02},
because of the rich phase behavior and the geophysical and
technological implications. It has been shown experimentally \cite{kooi:01}, theoretically
\cite{wens:01}, and by simulation \cite{bate:99}, that polydispersity
in the size of the platelets strongly affects the phase behavior.
Whereas the theoretical studies have focused on the understanding of the
interactions and the  phase behavior of homogeneous bulk fluids, 
experimentally it turns out that boundaries such as the walls of the sample cells 
have a pronounced influence on  the phase behavior \cite{brow:98,brow:99,kooi:01}.
Liquid-liquid or wall-liquid interfaces are intrinsic inhomogeneities 
of the experimental samples which have been studied recently 
\cite{brow:98,kooi:98,brow:99,kooi:01,kroo:01a}.

Here we study inhomogeneous colloidal fluids consisting of
non-spherical particles by examining both binary hard-platelet and
binary hard-rod fluids within the Zwanzig model \cite{zwan:63}. Platelets
or rods are represented by square parallelepipeds and the allowed orientations
of the normal of the particles along their main axis of symmetry are restricted
to three mutually perpendicular directions, rather than a continuous range
of orientations in space (see Fig.~\ref{fig1}). Zwanzig's model may be considered
as a coarse-grained version of the Onsager model which allows for continuously varying
orientations \cite{onsa:49}. Zwanzig's model offers the advantage
that the difficult determination of inhomogeneous density profiles becomes 
numerically straightforward, allowing one to study various aspects of inhomogeneous 
binary hard-platelet and binary hard-rod fluids in detail.
On the basis of recent experience with monodisperse hard-rod 
fluids \cite{roij:00a,roij:00b,dijk:01} the Zwanzig model is expected to
provide a qualitatively correct description of the aforementioned colloidal 
suspensions by focusing on the entropic properties. In studying both binary platelet 
fluids and binary rod fluids we address the problem of a possible surface transition 
from uniaxial to biaxial symmetry. To the best of our knowledge properties of 
binary rod fluids near a hard wall or in a  slit pore have also not been studied 
before. Our study provides a direct comparison of the structural properties 
and of the behavior of fluids consisting of rodlike and platelike particles, 
respectively.

In Sec. II we describe the density-functional theory
and the third-order virial  excess free energy functional.
Section III presents bulk phase diagrams of binary mixtures of thin platelets
and binary mixtures of thin rods, showing how the density gap at the
isotropic-nematic transition varies with the mole fraction of the larger particles.
In Sec. IV we determine the density and orientational profiles as well as the
excess adsorptions of the fluids near a hard wall. The calculations exhibit a
wall-induced surface transition from uniaxial to biaxial symmetry for fluids
consisting of rods. Binary hard-rod and binary hard-platelet fluids confined by two
parallel hard walls are investigated in Sec. V. For sufficiently large slit widths
a first-order capillary nematization transition is found. Our results are summarized
in Sec. VI.

\section{Density functional for the Zwanzig model}
We consider a binary mixture of hard rectangular particles of size 
$L_i\times D_i\times D_i$ ($i=1,2$) \cite{zwan:63}.
The number density of the centers of mass of the particles of species 
$i$ at a point ${\bf r}$ is denoted by $\rho^{(i)}_\beta({\bf r})$.
The number density of the centers of mass of the particles of species 
$i$ at a point ${\bf r}$ is denoted by $\rho^{(i)}_\beta({\bf r})$. 
The position of the center of mass ${\bf r}$ 
is continuous, while the allowed orientations of
\begin{figure}
\vspace*{-1.4cm}
\hspace*{-2.8cm} 
\includegraphics[width=0.75\linewidth]{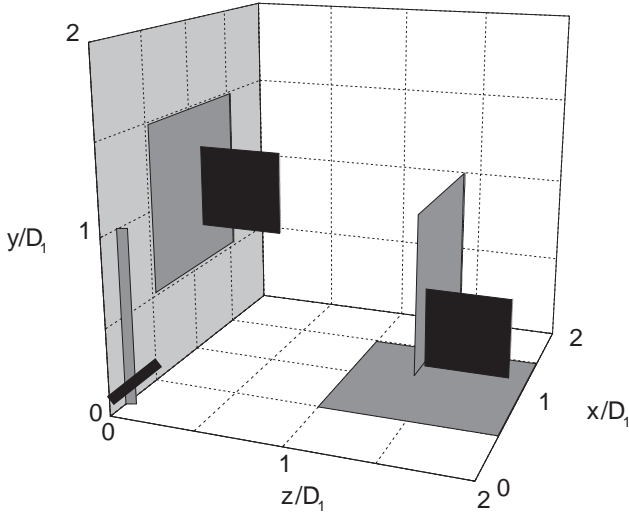}
\caption{The system under consideration consists of a binary fluid of thin
platelets  of surface sizes $D_1\times D_1$ (gray squares) and 
$D_2\times D_2$ (black squares) in contact with a planar hard wall 
at $z=0$.  The figure displays  configurations 
contributing to the second order virial term $\rho_x^{(2)}(z_1)\rho_z^{(1)}(z_2)$ 
[left], and the third order virial term 
$\rho_x^{(2)}(z_1)\rho_y^{(1)}(z_2)\rho_z^{(1)}(z_3)$ [right] for $D_1=2D_2$.
For comparison a configuration contributing to the second order virial term 
$\rho_x^{(2)}(z_1)\rho_y^{(1)}(z_2)$ of a binary rod fluid is shown.
Within this model of only three discrete orientations
particles lying very close to the wall must adopt a fully parallel alignment.}
\label{fig1} 
\end{figure}
the normal of the particles 
along their main axis of symmetry is restricted to directions $\beta=x,y,z$.  
The equilibrium density profiles of the mixture under the influence of external 
potentials $V^{(i)}_\beta({\bf r})$ minimize the grand potential functional
\begin{eqnarray} \label{eq1}
\Omega[\rho^{(i)}_\beta({\bf r})]&\!\!=\!\!&\sum_{i=1}^2\sum_{\beta}\int d{\bf r}\,
\rho^{(i)}_\beta({\bf r})
\left[k_BT\left(\ln[\Lambda_i^3\rho^{(i)}_\beta({\bf r})]\!-\!1\right)\right.
\nonumber
\\&&-\left.\mu_i+ V^{(i)}_\beta({\bf r})\right]+F_{ex}[\rho^{(i)}_\beta({\bf r})]\,,
\end{eqnarray}
where $\Lambda_i$ are the thermal de Broglie wavelengths and $\mu_i$ are the 
chemical potentials. Within a third order virial approximation the 
excess free energy functional $F_{ex}[\rho^{(i)}_\beta({\bf r})]$ is given by
\begin{eqnarray} \label{eq2}
\lefteqn{F_{ex}[\rho^{(i)}_\beta({\bf r})]=}\nonumber
\\-\!\!\!\!\!\!&&\frac{k_BT}{2}\sum_{i,j=1}^2
\sum_{\beta_1,\beta_2}\!\!\int\!\! d{\bf r}_1\,d{\bf r}_2\,
f^{(i,j)}_{\beta_1,\beta_2}({\bf r}_1,{\bf r}_2)
\rho^{(i)}_{\beta_1}({\bf r}_1)\rho^{(j)}_{\beta_2}({\bf r}_2)\nonumber
\\\times\!\!\!\!\!\!&&\!\left[1\!+\!\frac{1}{3}\!\sum_{k=1}^2
\!\sum_{\beta_3}\!\!\int\!\! d{\bf r}_3\,
f^{(j,k)}_{\beta_2,\beta_3}({\bf r}_2,{\bf r}_3)
f^{(k,i)}_{\beta_3,\beta_1}({\bf r}_3,{\bf r}_1)
\rho^{(k)}_{\beta_3}({\bf r}_3)\right]\!,\nonumber
\\&&
\end{eqnarray}
where $f^{(i,j)}_{\beta_1,\beta_2}({\bf r}_1,{\bf r}_2)$ is the Mayer 
function. The Mayer function equals $-1$ if the particles overlap and is zero 
other\-wise. With the definition 
\mbox{$S_{\alpha,\beta}^{(i)}=D_i+(L_i-D_i)\delta_{\alpha,\beta}$}, 
which represents the spatial extent in direction $\alpha\!=\!x,y,z$ of a particle
with orientation $\beta$ of the normal, the Mayer function can be written 
explicitly as 
\begin{eqnarray} \label{eq3}
\lefteqn{f^{(i,j)}_{\beta_1,\beta_2}({\bf r}_1,{\bf r}_2)=}\nonumber
\\&&-\prod_{\alpha=1}^3
\Theta\left(\frac{1}{2}\left(S_{\alpha,\beta_1}^{(i)}+S_{\alpha,\beta_2}^{(j)}
\right)-|r_{\alpha,1}-r_{\alpha,2}|\right)\,,
\end{eqnarray}
where $r_{\alpha,1}$ is the projection of the position vector ${\bf r}_1$ in 
$\alpha$ direction and $\Theta(r)$ is the Heaviside step function. The density 
functional theory is completely specified by the excess free energy functional 
and the Mayer function. The necessity for including the third order density term 
in Eq.~(\ref{eq2}), which is not present 
in the Onsager second virial approximation \cite{onsa:49} used in the description 
of thin rods, already follows from recent calculations of equilibrium properties of 
a homogeneous fluid consisting of monodisperse thin platelets \cite{harn:01a}.

For model systems of hard particles near a structureless wall at $z=0$, apart from 
a possible surface freezing at high densities, nonuniformities of the density occur 
only in the $z$ direction, so that $\rho^{(i)}_{\beta}({\bf r})=\rho^{(i)}_{\beta}(z)$. 
Hence the excess free energy functional can be written as
\begin{eqnarray} \label{eq4}
\lefteqn{F_{ex}[\rho^{(i)}_\beta(z)]=}\nonumber
\\-\!\!\!\!\!&&\frac{k_BT}{2}\sum_{i,j=1}^2
\sum_{\beta_1,\beta_2}\int dz_1\,dz_2\,
l^{(i,j)}_{\beta_1,\beta_2}(z_1,z_2)
\rho^{(i)}_{\beta_1}(z_1)\rho^{(j)}_{\beta_2}(z_2)\nonumber
\\\times\!\!\!\!\!\!&&\!\left[1\!+\!\frac{1}{3A}\!\sum_{k=1}^2
\!\sum_{\beta_3}\!\int\! dz_3\,
l^{(j,k)}_{\beta_2,\beta_3}(z_2,z_3)
l^{(k,i)}_{\beta_3,\beta_1}(z_3,z_1)
\rho^{(k)}_{\beta_3}(z_3)\right]\!\nonumber
\\&&
\end{eqnarray}
with
\begin{eqnarray} \label{eq5}
l^{(i,j)}_{\beta_1,\beta_2}(z_1,z_2)&=&\int dx_1\,dy_1\, dx_2\,dy_2\,
f^{(i,j)}_{\beta_1,\beta_2}({\bf r}_1,{\bf r}_2)\nonumber
\\&&\hspace{-1.2cm}=-A\left(S_{x,\beta_1}^{(i)}+S_{x,\beta_2}^{(j)}\right)
\left(S_{y,\beta_1}^{(i)}+S_{y,\beta_2}^{(j)}\right)\nonumber
\\&&\hspace{-0.8cm}\times\Theta\left(\frac{1}{2}
\left(S_{z,\beta_1}^{(i)}+S_{z,\beta_2}^{(j)}
\right)-|z_1-z_2|\right)\,,
\end{eqnarray}
where $A$ is the macroscopic surface area in the $x-y$ plane.
The particular factorization of the Mayer function (\ref{eq3}), which 
results from both particle shape and restricted orientations, leads
to the relative simplicity of the functions
$l^{(i,j)}_{\beta_1,\beta_2}(z_1,z_2)$. 
Figure \ref{fig1} displays a second and third order virial 
contribution schematically. 

Third order virial contributions for thin 
rods ($D_i\ll L_j$, $i,j=1,2$) are negligible due to the small intermolecular 
interaction between rods \cite{zwan:63,onsa:49}. For thin platelets 
($L_i\ll D_j$, $i,j=1,2$) the truncation of the virial expansion after the 
second order cannot be justified 
because of the nonzero probability of intersection even at small thickness
$L_i$ (see Fig.~\ref{fig1}).

\section{Isotropic and nematic bulk phases}
Based on the density functional in Eqs.~(\ref{eq1}), (\ref{eq4}), and (\ref{eq5}) 
we study first the homogeneous bulk fluid with $V^{(i)}_\beta({\bf r})=0$ 
in a macroscopic volume $V$. The equilibrium profiles are then constant
($\rho_\beta^{(i)}({\bf r})=\rho_\beta^{(i)}$) and the Euler-Lagrange equations
resulting from the stationarity conditions 
$\partial \Omega[\rho_\beta^{(i)}]/\partial\rho_\beta^{(i)}=0$ 
for the binary platelet mixture read
($\beta_1\neq\beta_2\neq\beta_3$, $i\neq j$):
\begin{eqnarray}  \label{eq6}
\lefteqn{\ln(\Lambda_i^3\rho_{\beta_1}^{(i)})=}\nonumber
\\&&\mu_i (k_BT)^{-1}-
2D_i^3\left[\rho_{\beta_2}^{(i)}+\rho_{\beta_3}^{(i)}\right]\nonumber
\\&&-D_iD_j(D_i+D_j)\left[\rho_{\beta_2}^{(j)}+\rho_{\beta_3}^{(j)}\right]\nonumber
-D_i^6\rho_{\beta_2}^{(i)}\rho_{\beta_3}^{(i)}
\\&&-D_i^2D_j^4\rho_{\beta_2}^{(j)}\rho_{\beta_3}^{(j)}
-D_i^4D_j^2\left[\rho_{\beta_2}^{(i)}\rho_{\beta_3}^{(j)}
+\rho_{\beta_2}^{(j)}\rho_{\beta_3}^{(i)}\right]\,.
\end{eqnarray}

The Euler-Lagrange equations
for a binary mixture of thin rectangular rods ($D=D_i=D_j$)
are given by \cite{clar:92}:
\begin{eqnarray}  \label{eq7}
\ln(\Lambda_i^3\rho_{\beta_1}^{(i)})&=&\mu_i (k_BT)^{-1}-
2L_i^2D\left[\rho_{\beta_2}^{(i)}+\rho_{\beta_3}^{(i)}\right]\nonumber
\\&&-2L_iL_jD\left[\rho_{\beta_2}^{(j)}+\rho_{\beta_3}^{(j)}\right]\,.
\end{eqnarray}
The Euler-Lagrange equations (\ref{eq6}) for the platelet mixture
are independent of the platelet thickness $L_i$ because the
integrals over Mayer functions in Eq.~(\ref{eq2}) are independent of
$L_i$ for thin platelets. For example, the integral over the Mayer function
of two particles of species $i$ orthogonal to each other is given by
\begin{eqnarray} \label{eq7a}
V^{(i,i)}_{x,z}=\int d{\bf r}_1\,f^{(i,i)}_{x,z}({\bf r}_1,0)=
2D_i(L_i+D_i)^2\,.
\end{eqnarray}
In the limit of thin platelets ($L_i\ll D_i$) the integral over the
Mayer function reduces to $V^{(i,i)}_{x,z}\approx 2D_i^3$, which is the
prefactor of the second term on the right side of Eq.~(\ref{eq6}).
In other words, thin platelets have an excluded volume
(the volume which is denied to a platelet by the condition that it must not
intersect another platelet), although they have a vanishing volume $L_iD^2_i$
(see Fig.~\ref{fig1}). On the other hand, the integrals over Mayer functions
and the excluded volume of thin rods depend on both  $L_i$ and  $D_i$:
$V^{(i,i)}_{x,z}\approx 2L_i^2D_i$.
We have solved Eqs.~(\ref{eq6}) and (\ref{eq7}) numerically 
for given chemical potentials. 
For convenience the total number density 
$\rho_b=\sum_i\sum_\beta\rho_{\beta}^{(i)}$ 
and the number densities $\rho_i$ of particles of species $i$ are introduced according to: 
$\rho_i=\rho_{x}^{(i)}+\rho_{y}^{(i)}+\rho_{z}^{(i)}$, $\rho_1+\rho_2=\rho_b$.
The theory has been formulated in a way that is completely symmetrical 
with respect to the three coordinate axis. Hence there must be a corresponding threefold 
degeneracy in the results. We define the $z$ axis as the 
preferred 
\begin{figure}
\vspace*{-1.0cm}
\includegraphics[width=1\linewidth]{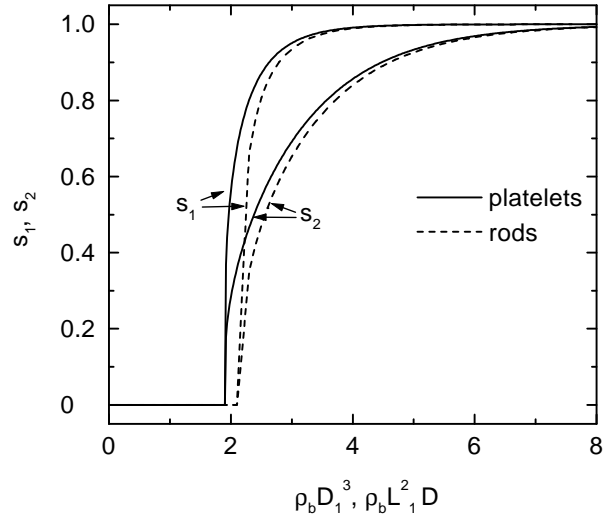}
\vspace*{-4.5cm}
\caption{Relative bulk nematic order parameters $s_1$ (upper curves) and  
$s_2$ (lower curves) for a binary mixture of thin rectangular platelets
(solid curves, $D_1=2^{2/3}D_2$) and thin rectangular rods
(dashed curves, $L_1=2L_2$, $L_1/D \to \infty$). The number densities of the larger
and the smaller particles are fixed to $\rho_1=\rho_2$. 
Isotropic orientations at low densities are characterized by $s_i=0$, 
while nematic ordering  $(0<s_i \le 1)$ is preferred for higher densities.} 
\label{fig2}   
\end{figure}  
coordinate axis and consider the relative uniaxial nematic order parameters 
$s_i=[\rho_{z}^{(i)}-(\rho_{x}^{(i)}+\rho_{y}^{(i)})/2]/\rho_i$.  
A typical set of ($s_1$, $s_2$) as a function of $\rho_b$, with $\rho_1=\rho_2$, 
$L_1=2L_2$ for rods and $D_1=2^{2/3}D_2$ for platelets is shown in Fig.~\ref{fig2}.
The size ratios of the rods and platelets have been fixed such that the 
second virial coefficients $b_2$ of the equation of state of 
the monodisperse fluids ($\rho_2=0$) in the isotropic phase are equal:
\begin{eqnarray}  \label{eq9}
\Omega&=&-\rho_b\left[1+b_2\rho_b+b_3\rho_b^2\right]k_BTV\,,
\end{eqnarray}
with $b_2=2L_1^2D/3$ for thin rods, and $b_2=2D_1^3/3$ for thin platelets.
For small values of $\rho_b$ the isotropic phase ($s_1=s_2=0$) is stable.
At a critical density another set of solutions, with $0<s_i \le 1$, appears
which represents the more favorable nematic phases. The isotropic-nematic (IN) 
transition of the binary platelet mixture takes place at a smaller density than 
for the rod mixture. We notice that the chemical potentials $\mu_i$ can be written 
as a function of $s_i$ and $\rho_i$ using Eqs.~(\ref{eq6}) and (\ref{eq7}).
At this point it is convenient to introduce the variables 
$\mu^\star_i=\mu_i-k_BT\ln(\Lambda_i^3/c_i)$, where $c_i=L_1^2D$ for the rod fluid 
and  $c_i=D_1^3$ for the platelet fluid. In the following numerical data
are given in terms of $\mu^\star_i$ and we drop the star in order to avoid a 
clumsy notation.
The compositions, densities, order parameters, and thermodynamic properties 
of the IN coexistence phases are found by solving the coexistence conditions:
$\mu_{iI}=\mu_{iN}$ and $p_I=p_N$
where $\mu_{iI}$, $\mu_{iN}$ and $p_I=-\Omega_I/V$, $p_N=-\Omega_N/V$ are the chemical 
potentials and the pressure of the isotropic and the nematic phase, respectively. 
The phase diagrams for
\begin{figure}
\vspace*{-0.6cm}
\includegraphics[width=1\linewidth]{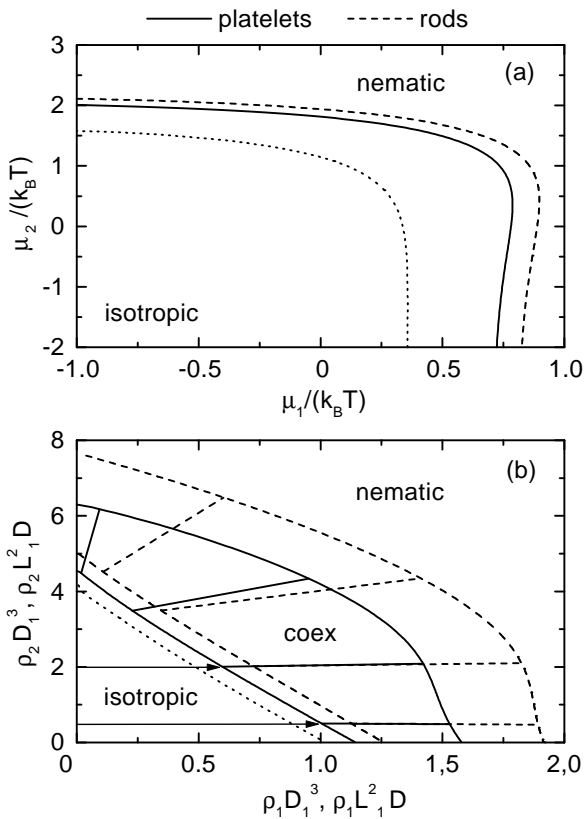}
\vspace*{-1cm}
\caption{(s) Bulk phase diagram of a fluid consisting of a binary mixture of thin
platelets (solid curve, $D_1=2^{2/3}D_2$) and a binary mixture of thin rods 
(dashed curve, $L_1=2L_2$, $L_1/D \to \infty$) as a function of the chemical
potentials $\mu_1$ and $\mu_2$. 
(b) Bulk phase diagram  of the same fluids in the density-density, i.e., $\rho_1-\rho_2$
plane (with the same line code as in (a)). The straight lines are tielines illustrating 
isotropic-nematic coexistence. In (a) and (b) the dotted lines mark the locations of the
uniaxial-biaxial transition densities of the binary rod fluid in contact with
a hard wall. In Figs.~\ref{fig5} and \ref{fig6} excess adsorptions 
near a hard wall are shown along the two thermodynamic paths indicated by arrows.}
\label{fig3}
\end{figure}
binary platelet and binary rod mixtures are displayed in
Fig.~\ref{fig3}. The calculations render the concentration in the isotropic phase
always to be less than in the nematic phase. 
For monodisperse platelet 
fluids the density gap at the 
IN transition $\triangle \rho=(\rho^{(N)}-\rho^{(I)})/\rho^{(I)}=0.23$ is 
smaller than the one for monodisperse rod fluids ($\triangle \rho=0.52$). 
The different size of the density gap is due to larger intermolecular interactions 
between platelets as compared with those between rods. Moreover, a widening of the 
IN coexistence region is observed at intermediate values of the mole fraction 
of the larger particles.  With increasing the size ratios of the particles 
($L_1=3L_2$, $D_1=3^{2/3}D_2$) the calculations exhibit a greater degree of 
fractionation between the two coexisting phases. 
The smaller particles are 
preferentially in the isotropic phase. The results for 
rods are in agreement with earlier calculations \cite{clar:92}.

\section{Hard-rod and hard-platelet fluids near a hard wall}
The density and orientational profiles of both components of binary 
hard-rod and binary hard-platelet mixtures
close to a planar hard wall are obtained by a numerical 
minimization of the grand potential functional 
(\ref{eq1}) with the excess free energy functional (\ref{eq4}).
The results are conveniently expressed in terms of the 
orientationally averaged number density profiles 
\begin{equation} \label{eq9a}
\rho_i(z)=\rho_{x}^{(i)}(z)+\rho_{y}^{(i)}(z)+\rho_{z}^{(i)}(z)\,,
\end{equation}
position-dependent nematic order parameters 
\begin{equation} \label{eq9b}
s_i(z)=\frac{\rho_{z}^{(i)}(z)-0.5[\rho_{x}^{(i)}(z)+\rho_{y}^{(i)}(z)]}{\rho_i(z)}\,,
\end{equation}
and position-dependent biaxial order parameters
\begin{equation} \label{eq9c}
q_i(z)=\frac{\rho_{x}^{(i)}(z)-\rho_{y}^{(i)}(z)}{\rho_i(z)}\,.
\end{equation}

At small distances from the wall the value of the nematic order parameter 
reflects the geometric constraints. A platelet [rod] lying very closely
to the wall must adopt a fully parallel alignment (see Fig.~\ref{fig1}),
so that the nematic order parameters reach their limiting values $s_i(0)=1$
[$s_i(0)=-1/2$] there, whereas the isotropic orientation $s_i(z)=0$
is attained at large
distances from the wall. Orientational profiles of 
biaxial symmetry are described by $s_i(z)\neq 0$, $q_i(z)\neq 0$, where a 
positive or negative sign of $q_i(z)$ signals a spontaneous preferential 
alignment of the normals parallel to the $x$-axis or $y$-axis, respectively.

\subsection{Monodisperse fluids}
\begin{figure}
\vspace*{-1.0cm}
\includegraphics[width=1\linewidth]{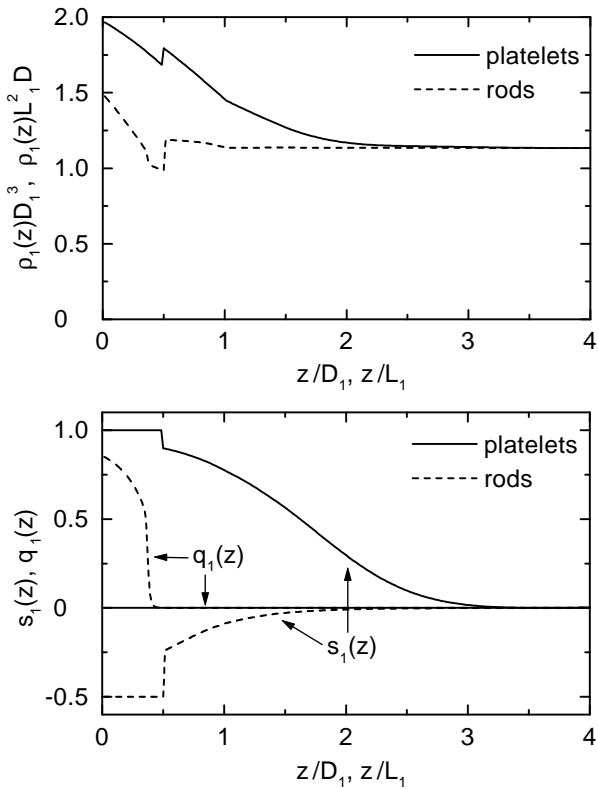}
\vspace*{-1cm}
\caption{Orientationally averaged number density profiles $\rho_1(z)$, 
nematic order parameters $s_1(z)$, and biaxial order parameters $q_1(z)$ for 
monodisperse thin platelets (solid curves) and thin rods (dashed curves, $D/L_1 \to \infty$) 
in contact with a planar hard wall at $z=0$. Positive [negative] values of the 
nematic order parameters indicate that the platelets [rods] are preferentially aligned 
parallel to the wall. Biaxial symmetry of rods near the wall is characterized by 
$q_1(z)\neq 0$. Here $q_1(z)\neq 0$ for $z>L_1/2$. The bulk density is fixed to 
$\rho_1D_1^3=1.13$ for the platelets and $\rho_1L_1^2D=1.13$ for the rods.}  \label{fig4}   
\end{figure}

The phase behavior of monodisperse hard-rod fluids ($\rho_2(z)=0$) near a 
structureless wall has been investigated using the Zwanzig model and a 
wall-induced transition from uniaxial to biaxial symmetry upon increasing the 
bulk density has been found \cite{roij:00a,roij:00b}. Moreover, we have recently 
investigated both monodisperse hard-platelet fluids and monodisperse hard-rod fluids 
near a structureless wall at low bulk densities using a model that allows for 
continuous orientations of the particles \cite{harn:02}. Here we extend these
previous calculations and present in Fig.~\ref{fig4} the calculated order parameters 
of monodisperse hard-platelet fluids and monodisperse hard-rod fluids for high 
bulk densities. The sharp cusps at $z=D_1/2$ 
for platelets, and at $z=L_1/2$ for rods, reflect the discontinuities of 
$\rho_{x}^{(1)}(z)$ and $\rho_{y}^{(1)}(z)$ for platelets, and of 
$\rho_{z}^{(1)}(z)$ for rods, which determine 
the value of the nematic order parameters close to the wall. The most noteworthy 
feature is that no biaxiality is found for the platelet fluid ($q_1(z)=0$)
while the loss of translational invariance due 
to the hard wall breaks the uniaxial symmetry of the rod fluid ($q_1(z)\neq 0$).
In order to study the possible onset 
of biaxiality, we rewrite the Euler-Lagrange equations in terms of the
orientationally averaged number density profile and of the order parameters:
\begin{eqnarray}  \label{eq10}
\lefteqn{\ln[\Lambda_1^3\rho_x^{(1)}(z)]-\ln[\Lambda_1^3\rho_y^{(1)}(z)]=}\nonumber
\\&&\ln\left[\frac{1-s_1(z)+\frac{3}{2}q_1(z)}{1-s_1(z)-\frac{3}{2}q_1(z)}\right]\equiv\triangle_1(z)\,,
\end{eqnarray}
with 
\begin{eqnarray}  \label{eq11}
\triangle_1(z)=2L_1^2D\rho_1(z)q_1(z)\,
\end{eqnarray}
for the hard-rod fluid and
\begin{eqnarray}  \label{eq12}
\triangle_1(z)&=&\frac{1}{D_1}\int\limits_{z-D_1}^{z+D_1}\!\!dz_1\,
D_1^3\rho_1(z_1)[q_1(z_1)\nonumber
\\&+&\!\!\!\!\!\!\frac{4}{D_1}
\int\limits_{z-\frac{D_1}{2}}^{z+\frac{D_1}{2}}\!\!dz_2\,
\Theta\left(\frac{D_1}{2}-|z_1-z_2|\right)q_1(z_2)]
\end{eqnarray}
for the hard-platelet fluid. One easily finds that the uniaxial distribution, 
with $q_1(z)=0$, is a solution of Eqs.~(\ref{eq11}) and (\ref{eq12}) for any
density $\rho_1(z)$ and nematic order profile $s_1(z)$. Biaxial distributions 
$q_1(z)\neq 0$ for the rod fluid are possible if $\rho_1(z)\ge \rho_1^{(UB)}(z)$, where 
the uniaxial-biaxial (UB) transition density $\rho_1^{(UB)}(z)$ follows from a 
low-$q_1$ expansion of  
$\triangle_1(z)=3q_1(z)/[1-s_1(z)]+O(q^3_1(z))$
for $s_1(z)\neq 1$: 
\begin{eqnarray}  \label{eq13}
\rho_1^{(UB)}(z)=\frac{3}{2[1-s_1(z)]L_1^2D}\,.
\end{eqnarray}
Since $s_1(z)=-1/2$ ($0\le z \le L_1/2$) is the minimum value of $s_1(z)$ 
for rods and because $\rho_1^{(UB)}(z)$ decreases with decreasing $s_1(z)$,
it follows that local biaxiality starts to develop if $\rho_1(z)=1$ in the 
interval $0\le z \le L_1/2$ (see Fig.~\ref{fig4}). For platelets the value 
$s_1(z)=1$ ($0\le z \le D_1/2$) is determined by the geometric constraint. 
Hence $q_1(z)=0$ is the only solution of Eqs.~(\ref{eq10}) and (\ref{eq12}) 
close to the wall.

\subsection{Binary fluids}
\begin{figure}
\vspace*{-1.0cm}
\includegraphics[width=1\linewidth]{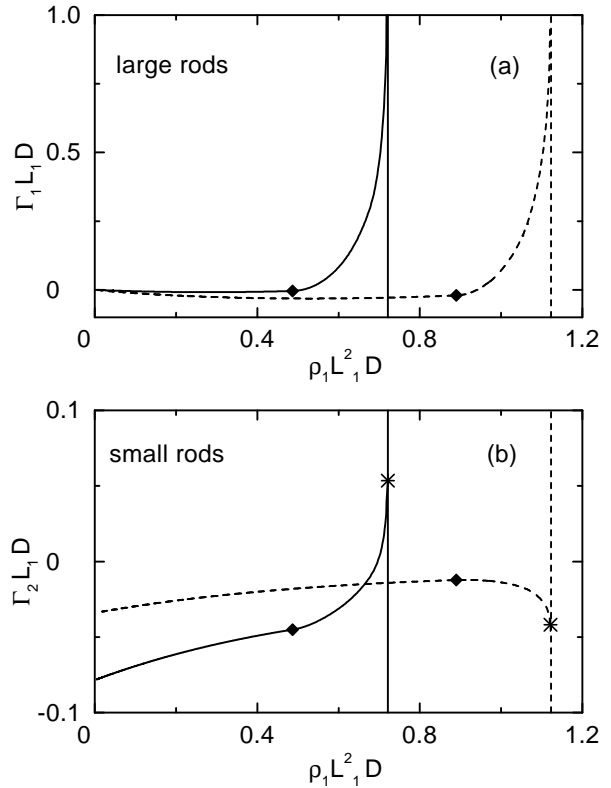}
\vspace*{-1cm}
\caption{The excess adsorptions $\Gamma_1$ and $\Gamma_2$ of large ((a)) and small 
((b)) rods, respectively, ($L_1=2L_2$, $L_1/D \to \infty$)
near a hard wall as a function of the bulk density $\rho_1$ of the large 
rods for two values of the bulk density of the small rods: 
$\rho_2L_1^2D=2$ (solid lines), $\rho_2L_1^2D=0.5$ (dashed lines).
The corresponding thermodynamic paths are indicated in
Fig.~\ref{fig3} (a).
The diamonds and the vertical lines mark the location of the uniaxial-biaxial
transition densities $\rho_1^{(UB)}$ and the densities of the isotropic
bulk phase at isotropic-nematic coexistence $\rho_1^{(I)}$, respectively. 
For comparison, $\rho_1^{(UB)}$ and $\rho_2$ are marked by the dotted line
and arrows in the bulk phase diagram displayed in Fig.~\ref{fig3} (a).
$\Gamma_1$ diverges logarithmically as $\rho_1\to \rho_1^{(I)}$, while 
$\Gamma_2$ attains finite values ($\ast$) via square-root cusp singularities.}
\label{fig5}
\end{figure}

The binary fluids considered in this subsection consist either of two types
of thin rods ($L_1=2L_2$) or of two types of thin platelets ($D_1=2^{2/3}D_2$).
We focus on the numerically determined excess adsorptions defined as 
\begin{eqnarray}  \label{eq14}
\Gamma_i&=&\int_0^\infty dz\,[\rho_i(z)-\rho_i]\,,
\end{eqnarray}
where $\rho_i=\rho_i(z\to\infty)$. Figures \ref{fig5} and \ref{fig6} display 
$\Gamma_i$ for binary rod fluids and binary platelet fluids, respectively. For a 
fixed bulk density of the small particles $\rho_2$ and for 
small bulk densities of the large particles $\rho_1$, the excess adsorption 
of the small particles 
increases upon increasing $\rho_1$. The reason for this is that the increasing
number of large particles lying close to the wall enforces the orientational
ordering of the small particles, leading to an 
enrichment of small particles near the wall because of reduced intermolecular 
interactions between the latter as compared to the isotropic bulk fluid.
Due to the same mechanism, $\Gamma_1$ increases upon increasing $\rho_2$
for constant bulk densities of the large particles $\rho_1$. The excess adsorption 
of the large particles exhibits a change of sign and sharp increase with 
increasing $\rho_1$ while a net depletion of the small particles is found 
for small bulk densities $\rho_2$. The calculation renders $\Gamma_1$ to
diverge logarithmically as $\rho_1\to \rho_1^{(I)}$, where $\rho_1^{(I)}$
is the bulk density of the large particles in the isotropic phase at the IN
transition (see lower curves in the upper figure of Fig.~\ref{fig3}).
Near $\rho_1^{(I)}$ the excess coverage can be fitted by
$\Gamma_1=A_1-A_2\ln(L_1^2D[\rho_1^{(I)}-\rho_1])$ for the rod fluid
and $\Gamma_1=B_1-B_2\ln(D_1^3[\rho_1^{(I)}-\rho_1])$ for the platelet fluid,
with fit parameters $A_1$, $A_2$ and $B_1$, $B_2$, while the excess coverage
of the small particles remains finite and attains
the critical value $\Gamma_2^{(c)}$
via a square-root cusp singularity
$\Gamma_2-\Gamma_2^{(c)} \sim \sqrt{\rho_1^{(I)}-\rho_1}$.
The logarithmic increase of $\Gamma_1$ is consistent with
complete wetting of the wall -- isotropic fluid interface by a nematic film
at $\rho_1^{(I)}$. Complete wetting is confirmed explicitly by the
vanishing of the contact angle (see Sec. IV).
One observes that the UB transition of the rods marks the onset 
of a pronounced variation of $\Gamma_1$, while $\Gamma_1$ increases more smoothly 
upon increasing $\rho_1$ for the binary platelet fluid, due to the absence of 
biaxiality. In order to study the possible onset of biaxiality for the binary 
rod fluid, we rewrite the Euler-Lagrange equations in terms of the density
profiles and order parameters:
\begin{figure}
\vspace*{-1.0cm}
\includegraphics[width=1\linewidth]{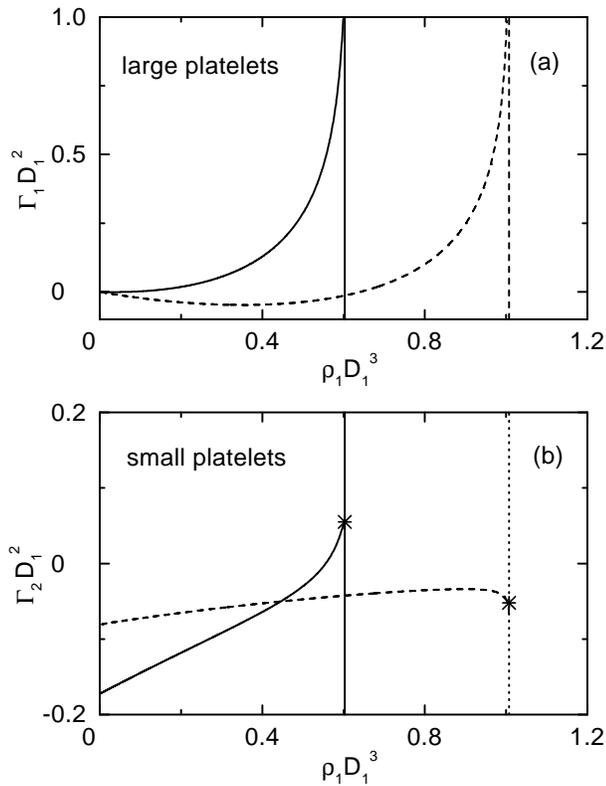}
\vspace*{-1cm}
\caption{The excess adsorptions $\Gamma_1$ and $\Gamma_2$ of large ((a)) and small ((b))
platelets, respectively, ($D_1=2^{2/3}D_2$) near a hard wall as a function of 
the bulk density $\rho_1$ of the large platelets for two values of the bulk 
density of the small platelets: $\rho_2D_1^3=2$ (solid lines), $\rho_2D_1^3=0.5$ 
(dashed lines). The corresponding thermodynamic paths are indicated in
Fig.~\ref{fig3} (a). The vertical lines mark the locations of the densities of the
isotropic phase at isotropic-nematic coexistence $\rho_1^{(I)}$. 
$\Gamma_1$ diverges logarithmically as $\rho_1\to \rho_1^{(I)}$, while
$\Gamma_2$ attains finite values ($\ast$) via square-root cusp singularities.} \label{fig6}
\end{figure}
\begin{eqnarray}  \label{eq15}
\lefteqn{\ln\left[\frac{1-s_1(z)+\frac{3}{2}q_1(z)}{1-s_1(z)-\frac{3}{2}q_1(z)}\right]=}\nonumber
\\&&\frac{L_1}{L_2}\ln\left[\frac{1-s_2(z)+\frac{3}{2}q_2(z)}{1-s_2(z)-
\frac{3}{2}q_2(z)}\right]\nonumber
\\&=&2D[L_1^2\rho_1(z)q_1(z)+2L_2^2\rho_2(z)q_2(z)]\,.
\end{eqnarray}
For any $s_i(z)$ and $\rho_i(z)$ there is a trivial uniaxial
solution $q_i(z)=0$ to Eq.~(\ref{eq15}).
Nontrivial biaxial solutions
$q_i(z)\neq 0$ exist at sufficiently high local densities
$\rho_i(z)\ge \rho_i^{(UB)}(z)$ with the UB transition densities
following from an expansion of the logarithms in Eq.~(\ref{eq15}):
\begin{eqnarray}  \label{eq16}
\lefteqn{\frac{3}{2[1-s_1(z)]}=}\nonumber
\\&&\rho_1^{(UB)}(z)L_1^2D+\frac{[1-s_2(z)]}{[1-s_1(z)]}
\rho_2^{(UB)}(z)L_2^2D\,.
\end{eqnarray}
In the case of a monodisperse hard-rod fluid  Eq.~(\ref{eq16}) reduces to 
Eq.~(\ref{eq13}). Since \mbox{$s_1(z)=s_2(z)=-1/2$} close to the wall,
local biaxiality sets in if $\rho_1(z)+L_2^2/L_1^2\rho_2(z)=1$ near the wall.
We notice that, independent of the size ratios of the rods,
biaxiality of one species of the fluid is always accompanied
by biaxiality of the other species, as expected on geometrical grounds.

\section{Binary hard-rod and hard-platelet fluids 
confined by two parallel hard walls}
The results of the previous section show that a hard wall favors planar 
nematic order (the main body of the particles is oriented parallel to the wall)
over isotropic order upon increasing the particle densities. We now consider
binary hard-rod and hard-platelet fluids confined by two parallel hard walls at 
$z=0$ and $z=H$ and investigate a possible capillary condensation of a 
nematic phase. Particularly, we calculate the surface contributions
defined via 
\begin{eqnarray} \label{eq17}
\Omega[\rho^{(i)}_\alpha(z)]&=&V\omega_b+2A\gamma_{wI}+A\omega(H)\,,
\end{eqnarray}
where $A$ is the area of a single surface, $\omega_b$ is the bulk grand-canonical 
potential density, and $V$ is defined as the volume of the container 
with its surface given by the position of the rim of the particles 
at closest approach so that $V=AH$. $\gamma_{wI}$ is the wall -- isotropic liquid 
surface tension in the absence of the second wall
and $\omega(H)$ is the finite size contribution.
We restrict our attention to chemical potentials $\mu_i$
smaller than the chemical potentials $\mu_i^{(IN)}$
at bulk isotropic-nematic coexistence.

\subsection{Surface tensions and wetting at a single hard wall}
\begin{figure}
\vspace*{-1.0cm}
\includegraphics[width=1\linewidth]{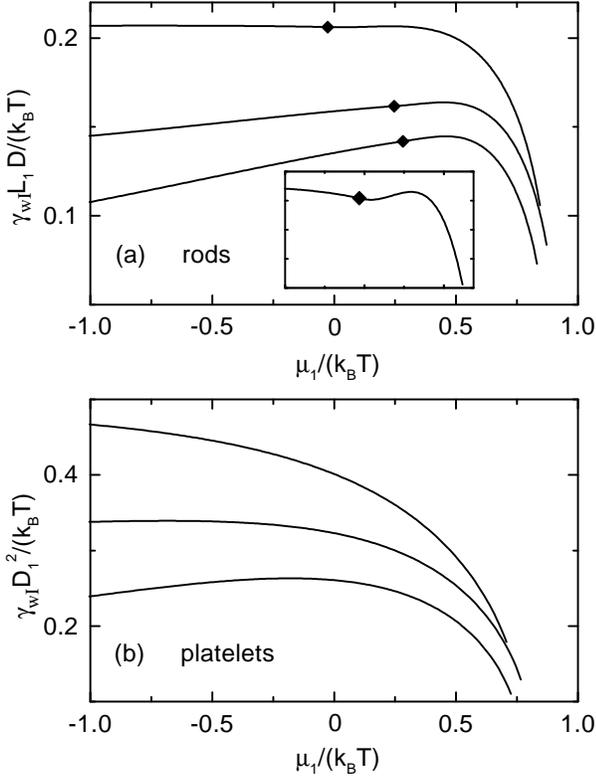}
\vspace*{-1cm}
\caption{The wall -- isotropic liquid surface tension $\gamma_{wI}$ of a binary hard-rod
fluid ($L_1=2L_2$, $L_1/D \to \infty$), (a), and a binary hard-platelet fluid
($D_1=2^{2/3}D_2$), (b), respectively. The chemical potential
of the small particles is kept fixed for each curve
and increases from bottom to top: $\mu_2/(k_BT)=-1, 0, 1$.
Bulk isotropic-nematic coexistence occurs at $\mu_1=\mu_1^{(IN)}(\mu_2)$, which
corresponds to the final points of the curves at the right
(see Fig.~\ref{fig3} (b)). The inset displays $\gamma_{wI}$ of the binary
hard-rod fluid for $\mu_2/(k_BT)=1$ with increased resolution.
The diamonds mark the uniaxial-biaxial transitions.}
\label{fig7}
\end{figure}

Figure \ref{fig7} displays the
surface tension $\gamma_{wI}$ as a function of the chemical potential
of the larger particles. The steric interaction between the particles,
which is more pronounced for the platelets, increases the surface tension
with increasing chemical potential. On the other hand,
nematic ordering of the particles, induced by the walls, leads to a decrease
of the surface tension for large chemical potentials. For fixed small chemical
potential $\mu_2$  of the small particles a maximum of $\gamma_{wI}$ as a
function of the chemical potential $\mu_1$ of the large particles
is observed.

For large negative chemical potentials, i.e., in the limit of
noninteracting particles, the wall -- isotropic liquid surface tension is given by
\begin{equation} \label{eq18}
\frac{\gamma_{wI}L_1D}{k_BT}=\frac{1}{2}\left[\exp\left(\frac{\mu_1}{k_BT}\right)+
\frac{L_2}{L_1}\exp\left(\frac{\mu_2}{k_BT}\right)\right]
\end{equation}
for the binary rod fluid, and
\begin{equation} \label{eq19}
\frac{\gamma_{wI}D_1^2}{k_BT}=\exp\left(\frac{\mu_1}{k_BT}\right)+
\frac{D_2}{D_1}\exp\left(\frac{\mu_2}{k_BT}\right)
\end{equation}
for the binary platelet fluid. The prefactor $1/2$ in Eq.~(\ref{eq18})
reflects the fact, that the orientationally averaged excluded volume due
to the wall is smaller for rods than for platelets.

For the binary rod mixture $\gamma_{wI}$ is a non-monotonic function of
$\mu_1$ close the uniaxial-biaxial transition.
The local minimum of $\gamma_{wI}$ displayed in the inset of
Fig.~\ref{fig7} disappears upon increasing or decreasing $\mu_2$,
i.e., in the limit of monodisperse fluids.

We have confirmed that complete wetting of the wall -- isotropic liquid interface 
by a nematic film occurs along the whole isotropic-nematic coexistence by 
observing a vanishing contact angle $\vartheta$:
\begin{equation} \label{eq18a}
\cos\vartheta=
\frac{\gamma_{wI}(\mu_i^{(IN)})-\gamma_{wN}(\mu_i^{(IN)})}{\gamma_{IN}}\,.
\end{equation}
Here $\gamma_{wI}(\mu_i^{(IN)})$ is the wall -- isotropic
liquid surface tension, $\gamma_{wN}(\mu_i^{(IN)})$
is the wall-nematic liquid surface tension, and $\gamma_{IN}$ is the
isotropic-nematic interfacial tension. All tensions are taken at
isotropic-nematic two-phase coexistence. The chemical potentials at the
IN transitions are denoted by $\mu_i^{(IN)}$
(see Fig.~\ref{fig3} (b)). 
\subsection{Film geometry}
\begin{figure}
\vspace*{-1.0cm}
\includegraphics[width=1\linewidth]{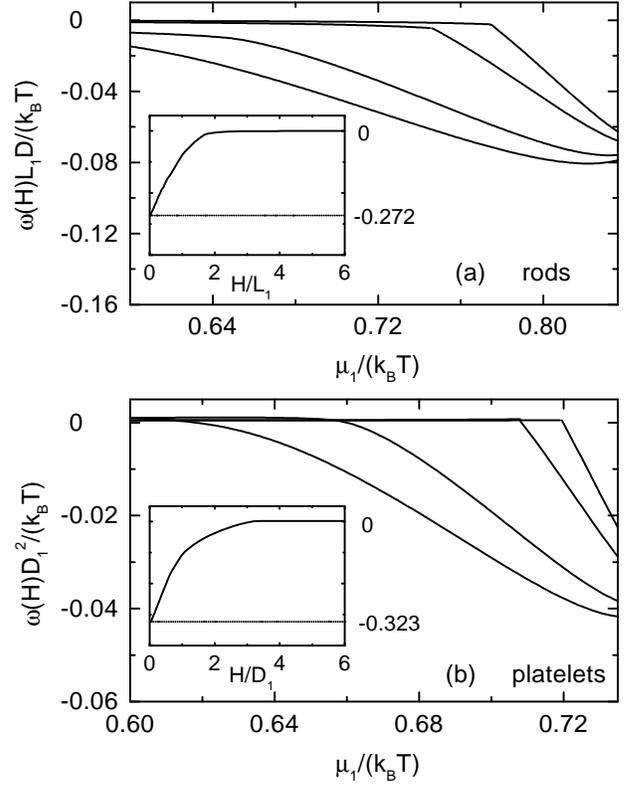}
\vspace*{-1cm}
\caption{The finite size contribution $\omega(H)$ of the grand potential
functional [see Eq.~(\ref{eq17})] of a binary hard-rod 
fluid ($L_1=2L_2$, $L_1/D \to \infty$), (a), and a binary hard-platelet fluid
($D_1=2^{2/3}D_2$), (b), respectively, confined in a slit of width $H$,
and in contact with an isotropic bulk reservoir at chemical potential $\mu_1$. 
The chemical potential of the small particles is kept fixed at
$\mu_2/(k_BT)=-1$. The  width of the slit increases from top to bottom:
$H/L_1=1.5, 1.75, 2.5 ,3$ in (a) and 
$H/D_1=3, 3.5, 5 ,6$ in (b). Bulk isotropic-nematic coexistence occurs at 
$\mu_1=\mu_1^{(IN)}(\mu_2)$, which are the maximum $\mu_1$-values displayed
(see Fig.~\ref{fig3} (b)). The insets display the solvation free energy 
$\omega(H)$ as function of $H$ at $\mu_1/(k_BT)=0.64$.}
\label{fig8}
\end{figure}

The results for the finite size contribution $\omega(H)$ are shown
in Fig.~\ref{fig8}. For sufficiently large slit widths $H$ the slope
of $\omega(H)$ as a function of the chemical potential of the large particles
changes discontinuously at a critical value,
signalling a first-order capillary nematization transition.
We emphasize that $\omega(H)$ is a smooth and monotonic
function of $\mu_1$ for fixed $H$ close to the uniaxial-biaxial transition
whereas as discussed above $\gamma_{wI}$ exhibits a local minimum for
binary rod mixtures close to the uniaxial-biaxial transition.
As function of $H$ the finite size contribution $\omega(H)$ corresponds 
to the solvation free energy for the immersed two plates acting as the 
confining walls for the fluid. According to the insets in Fig.~\ref{fig8}, 
$\omega(H)$ exhibits a minimum at $H=0$ (by construction,
$\omega(0)=-2\gamma_{wI}$ and $\omega(\infty)=0$) so that the solvation force 
$-d \omega(H)/dH$ is attractive.
For a  discussion on the repercussions 
of this solvation force in the context of colloidal stability we refer
to, e.g., Ref.~\cite{mao:95}. Whereas for the present systems 
$\gamma_{wI}>0$ (see Fig.~\ref{fig7}), for simple fluids the surface tension 
$\gamma_{wl}$ is typically negative whenever the wall-liquid attractions 
dominate over liquid-liquid attractions and even for hard walls
$\gamma_{wl}<0$ at densities well above liquid-vapor coexistence
for which drying does not occur. In those situations 
$\omega(0)>0$ and the solvation potential decreases as $H$ increases.

\begin{figure}
\vspace*{-1.0cm}
\includegraphics[width=1\linewidth]{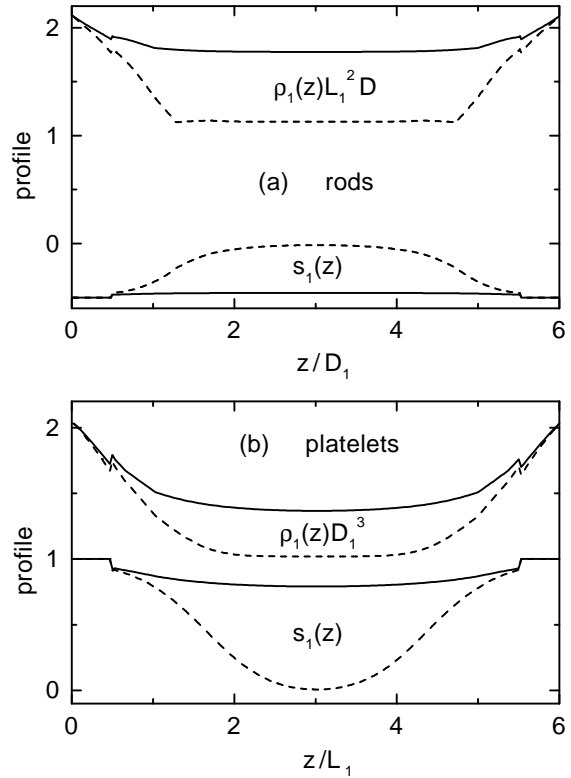}
\vspace*{-1cm}
\caption{Coexisting isotropic and capillary condensed nematic
profiles for a binary hard-rod fluid ($L_1=2L_2$, $L_1/D \to \infty$),
(a), and a binary hard-platelet fluid ($D_1=2^{2/3}D_2$), (b),
confined in a slit of width $H=6L_1$ and $H=6D_1$, respectively. 
The isotropic profiles of the large particles (dashed lines)
are characterized by $s_1(z)=0$ in the central region, while negative 
[positive] values of the nematic order parameters $s_1(z)$ close to 
the confining walls at $z=0$ and $z=6L_1$ [$z=6D_1$] indicate that 
the rods [platelets] are aligned parallel to the walls. The capillary 
condensed nematic phase (solid lines) is characterized by strong 
orientational ordering ($s_1(z)\neq 0$) throughout the slit. The chemical 
potential of the small particles is $\mu_2/(k_BT)=-1.0$ so that in the 
bulk $\rho_2^{(I)}L_1^2D=0.44$ and $\rho_1^{(I)}L_1^2D=1.14$ 
in (a) and $\rho_2^{(I)}D_1^3=0.44$ and $\rho_1^{(I)}D_1^3=1.02$ in (b).} 
\label{fig9}
\end{figure}

The occurrence of the capillary condensed
nematic phase can be inferred more directly from the density
and nematic order profiles displayed in Fig.~\ref{fig9}.
The density profile of the capillary condensed nematic phase is characterized by
a nematic phase throughout the slit, whereas the density profile of
the coexisting  phase decays toward an isotropic phase in the
middle of the slit. As expected from the presence of purely repulsive
walls, the total local midplane density $\rho_1(H/2)+\rho_2(H/2)$ of the
capillary condensed nematic phase is slightly smaller than the coexisting
nematic bulk density $\rho_1^{(N)}+\rho_2^{(N)}$.
For the binary rod fluid both the capillary condensed nematic phase and the
nematic film in the coexisting isotropic phase are biaxially symmetric.
The location of the uniaxial-biaxial transition has practically
not been altered by the confinement.

It is apparent from Fig.~\ref{fig9} that the interfacial
profile in the isotropic phase is larger for the platelet fluid than for the rod
fluid because of the relatively smaller intermolecular interactions
between rods as compared with those between platelets. We observe
coexistence between the isotropic and the capillary condensed nematic
phase provided $H\ge H_c(\mu_1,\mu_2)$. For sufficiently narrow slits
($H<H_c(\mu_1,\mu_2)$) a sharp capillary nematization transition no
longer occurs and is replaced by a steep but continuous filling. Hence
the capillary nematization transition ends in a capillary critical
point at a critical wall separation $H_c(\mu_1,\mu_2)$.

\begin{figure}
\vspace*{-1.0cm}
\includegraphics[width=1\linewidth]{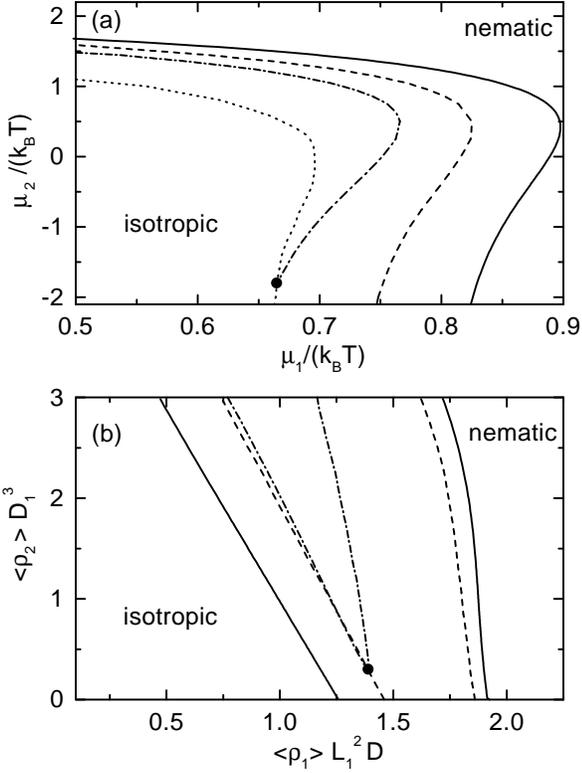}
\vspace*{-1cm}
\caption{(a)
Phase diagram of a confined  binary hard-rod fluid ($L_1=2L_2$, $L_1/D \to \infty$
as a function of the chemical potentials $\mu_1$, $\mu_2$, and wall separation $H$.
The solid line represents the bulk
phase diagram (see Fig.~\ref{fig3} (a)), while the dashed and the dash-dotted lines
correspond to wall separations  $H=6L_1$  and $H=2L_1$. For fixed $H<H_c(\mu_1,\mu_2)$ 
the capillary condensation transition is first order and terminates at a critical point. 
These critical points for various slit widths form the dotted line. One example of
such a capillary critical point is indicated by the solid circle.
There the corresponding line of first-order capillary transitions for
$H=2L_1$ (dash-dotted line) ends.
(b) Phase diagram of the same fluid plotted (with the same line code
as in (a)) as function of the
average number densities $\left\langle \rho_1 \right\rangle$ and
$\left\langle \rho_2 \right\rangle$ in the slit. In between corresponding
lines there is two-phase coexistence between isotropic and capillary condensed 
nematic phases. For a small slit width
$H=2L_1$ (dash-dotted line) the branches of the coexisting capillary
condensed nematic phase and isotropic phase end at a critical point
denoted by the solid circle.}
\label{fig10}
\end{figure}

\begin{figure}
\vspace*{-1.0cm}
\includegraphics[width=1\linewidth]{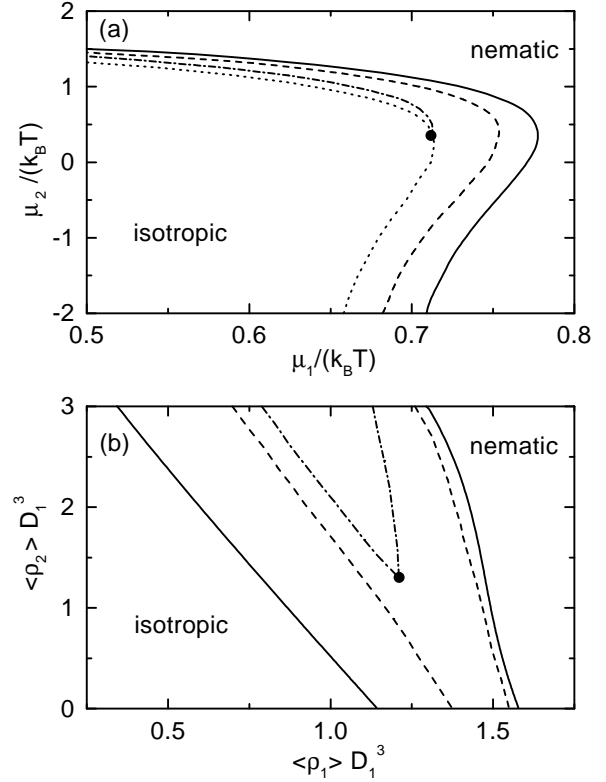}
\vspace*{-1cm}
\caption{
Phase diagrams of a confined  binary hard-platelet fluid ($D_1=2^{2/3}D_2$). 
The solid lines represent the bulk phase diagram (see Fig.~\ref{fig3}), 
while the dashed and the dash-dotted lines correspond to wall separations 
$H=6D_1$  and $H=2D_1$.
The critical points for various slit widths form the dotted line in (a).}
\label{fig11}
\end{figure}

We have determined the capillary nematization transition for various slit
widths and the phase diagrams for
binary rod and binary platelet mixtures constructed 
as function of the chemical potentials and the slit widths are shown in
Figs.~\ref{fig10} (a) and  \ref{fig11} (a).
Upon decreasing  the slit width ($H\ge H_c(\mu_1,\mu_2)$) the capillary 
nematization transition
is shifted to smaller chemical potentials reminiscent of the
shift of the capillary condensation transition in confined simple liquids.
Upon increasing the chemical potential $\mu_2$ of the small
particles the critical wall separation  decreases because the small particles
are preferentially in the isotropic phase in the 
middle of the slit. This leads to a depletion
of the large particles in the middle of the slit and prevents
a continuous filling of the slit even for rather small slit widths.

Figures \ref{fig10} (b) and \ref{fig11} (b) display an alternative  representation
of the capillary phase diagrams in terms of average number densities defined as
\begin{equation} \label{eq20}
\left\langle \rho_i \right\rangle=\frac{1}{H}\int\limits_0^Hdz\, \rho_i(z)\,.
\end{equation}
Here the curves with low values of $\left\langle \rho_i \right\rangle$
represent the coexisting isotropic phase and the curves with high
values of $\left\langle \rho_i \right\rangle$ the capillary condensed nematic phase.
Upon approaching the critical point  
($H<H_c(\left\langle \rho_1 \right\rangle,\left\langle \rho_2 \right\rangle)$) 
the difference of the average densities in the 
coexisting phases becomes smaller and vanishes at the critical point.
The shape of the coexistence curves in Figs.~\ref{fig10} (b) and \ref{fig11} (b)
reflects the fact that the difference in 
$\left\langle \rho_i \right\rangle$ between the two phases is used as 
order parameter. Our calculations lead to mean-field order 
parameter exponents; hence the dash-dotted lines in 
\mbox{Figs.~\ref{fig10} (b) and \ref{fig11} (b)} 
do not intersect under an acute angle. The actual rounding becomes visible 
only at higher resolutions.

\section{Summary}
We have applied a density-functional theory to fluids consisting of binary
hard-platelets and hard-rods near a single hard wall or confined in a slit
pore of width $H$. The particles are square parallelepipeds
with orientations restricted
to three mutually perpendicular directions (Fig.~\ref{fig1}).
The rectangular shape of the particles and their restricted
orientations allow one, within the framework of a third-order virial
approximation of the excess free energy functional, to determine
numerically the density profiles, orientational profiles, surface 
and finite-size contributions to the grand potential, and phase diagrams 
with the following main results.

(1) Figures \ref{fig2} and \ref{fig3} demonstrate that the bulk
isotropic-nematic transition of the binary platelet mixture
(size ratio: $D_1=2^{2/3}D_2$, where $D_i \times D_i$ is the surface
size of the thin platelets of species $i$) occurs at a smaller density
than for the rod mixture (size ratio: $L_1=2L_2$, where $L_i$ is the
length of the thin rods of species $i$). Moreover, the density gap
at the isotropic-nematic transition is smaller for the rod fluid than
for the platelet fluid because of larger intermolecular interactions
between platelets as compared with those of rods.

(2) Platelets [rods] lying very closely to a  planar hard wall must
adopt a fully parallel alignment (see Fig.~\ref{fig1}), so that the
nematic order parameters (Eq.~(\ref{eq9b})) reach their limiting 
values $s_i(0)=1$ [$s_i(0)=-1/2$] there (Fig.~\ref{fig4}). 
The biaxial order profiles (Eq.~(\ref{eq9c}))
for the inhomogeneous fluids at densities slightly below the bulk 
isotropic-nematic transition densities demonstrate biaxial symmetry 
($q_i(0)\neq 0$) of rods and uniaxial symmetry ($q_i(0)= 0$) of 
platelets near the wall (Fig.~\ref{fig4}).

(3) The excess adsorption $\Gamma_1$ of the large particles diverges
logarithmically as $\rho_1\to \rho_1^{(I)}$, where $\rho_1^{(I)}$ is the
bulk density of the large particles in the isotropic phase at the
isotropic-nematic transition (Figs.~\ref{fig5} and \ref{fig6}).
Complete wetting of the wall -- isotropic liquid interface by a nematic 
film is confirmed by observing a vanishing contact angle.

(4) The steric interaction between the particles increases the
wall -- isotropic liquid surface tension $\gamma_{wI}$ with increasing chemical
potential, whereas nematic ordering of the particles, induced by the walls,
leads to a decrease of the surface tension for large chemical potentials
(Fig.~\ref{fig7}).

(5) For sufficiently large slit widths the slope of the finite size
contribution to the grand potential as a function of the chemical potential
of the large particles changes discontinuously at a critical value,
signaling a first-order capillary nematization transition (Fig.~\ref{fig8}).
The density profile of the capillary condensed nematic phase is characterized
by a nematic phase throughout the slit, whereas the density profile of
the coexisting  phase decays toward an isotropic phase in the
middle of the slit (Fig.~\ref{fig9}). The isotropic-nematic interfacial 
profile is broader for the platelet fluid than for the rod fluid.

(6) Coexistence between the isotropic and the capillary condensed nematic
phase is observed provided the slit width $H$ is sufficiently large:
$H\ge H_c(\mu_1,\mu_2)$. For sufficiently narrow slits
($H<H_c(\mu_1,\mu_2)$) a sharp capillary nematization transition no
longer occurs and is replaced by a steep but continuous filling 
(Figs.~\ref{fig10} and  \ref{fig11}).


\begin{thebibliography}{99}
\bibitem{mait:00} See e.g., G. C. Maitland, Current Opinion in Colloid
and Interface Science {\bf 5}, 301 (2000).
\bibitem{brow:98} A. B. D. Brown, S. M. Clarke, and A. R. Rennie, Langmuir,
{\bf 14}, 3129 (1998).
\bibitem{kooi:98} F. M. van der Kooij and H. N. W. Lekkerkerker,
J. Phys. Chem. B {\bf 102}, 7829 (1998).
\bibitem{brow:99} A. B. D. Brown, C. Ferrero, T. Narayanan, and A. R. Rennie,
Eur. Phys. J. B {\bf 11}, 481 (1999).
\bibitem{kooi:01} F. M. van der Kooij and H. N. W. Lekkerkerker,
Phil. Trans. R. Soc. Lond. A {\bf 359}, 985, (2001).
\bibitem{kroo:01a} F. M. van der Kooij, D. van der Beek, and H. N. W. Lekkerkerker,
J. Phys. Chem. B, {\bf 105}, 1696, (2001).
\bibitem{bate:99} M. A. Bates and D. Frenkel,
J. Chem. Phys. {\bf 110}, 6553, (1999).
\bibitem{gali:00} A. Galindo, G. Jackson, and D. J. Photinos,
Chem. Phys. Lett., {\bf 325}, 631, (2000).
\bibitem{harn:01a} L. Harnau, D. Costa, and J.-P. Hansen, Europhys. Lett. 
{\bf 53}, 729 (2001).
\bibitem{wens:01} H. H. Wensink, G. J. Voerge, and H. N. W. Lekkerkerker,
J. Phys. Chem. B, {\bf 105}, 10610, (2001).
\bibitem{wens:02} H. H. Wensink, G. J. Voerge,
Phys. Rev. E, {\bf 65}, 031716, (2002).
\bibitem{harn:02} L. Harnau and S. Dietrich, Phys. Rev. E, 
{\bf 65}, 021505 (2002).
\bibitem{zwan:63} R. Zwanzig, J. Chem. Phys. {\bf 39}, 1714 (1963).
\bibitem{onsa:49} L. Onsager, Ann. (N.Y.) Acad. Sci. 
{\bf 51}, 627 (1949).
\bibitem{roij:00a} R. van Roij, M. Dijkstra, and R. Evans, Europhys. Lett. 
{\bf 49}, 350 (2000).
\bibitem{roij:00b} R. van Roij, M. Dijkstra, and R. Evans, J. Chem. Phys. 
{\bf 113}, 7689 (2000).
\bibitem{dijk:01} M. Dijkstra, R. van Roij, and R. Evans, Phys. Rev. E 
{\bf 63}, 051703 (2001).
\bibitem{clar:92} N. Clarke and T. C. B. McLeish, J. Phys. II France,
{\bf 2}, 1841 (1992).
\bibitem{mao:95} Y. Mao, M. E. Cates and H. N. W. Lekkerkerker, Phys. Rev. Lett.
{\bf 75}, 4548 (1995).

 
\end{thebibliography}
\end{document}